# Wideband resonant polarizers made with ultra-sparse dielectric nanowire grids

Jae Woong Yoon, Kyu Jin Lee, and Robert Magnusson *

*Department of Electrical Engineering, University of Texas at Arlington, Box 19016, Arlington, TX 76019, United States*

**Corresponding author*: magnusson@uta.edu

**Polarizers are essential in diverse photonics applications including display [1], microscopy [2], polarimetric astrophysical observation [3], laser machining [4], and quantum information processing [5]. Whereas conventional polarizers based on natural crystals and multilayer thin films are commonplace, nanostructured polarizers offer compact integrability [6,7], thermal stability in high-power systems [4,8], and space-variant vector beam generation [9,10]. Here, we introduce a new class of reflectors and polarizers fashioned with dielectric nanowire grids that are mostly empty space. It is fundamentally extremely significant that the wideband spectral expressions presented can be generated in these minimal systems. We provide computed results predicting high reflection and attendant polarization extinction in multiple spectral regions. Experimental results with Si nanowire arrays show ~200-nm-wide band of total reflection for one polarization state and free transmission for the orthogonal state. These results agree quantitatively with theoretical predictions.**

Subwavelength periodic thin-film polarizers fashioned in dielectric media provide robust high-power performance and feasible fabrication in most practical frequency domains. Innovative polarizers have been implemented by combining multilayer films with linear subwavelength gratings to induce polarization selectivity at normal incidence [11,12]. Advanced polarizer designs in simple architecture were subsequently demonstrated engaging guided-mode resonance effects [13-15]. These devices operate with a broadband resonant reflection in one polarization state and concomitant transmission in the orthogonal state. It is widely assumed that large refractive-index contrast and high average refractive index are necessary to support broadband performance with attendant multi-mode resonance excitation. In contrast, here we show that simple dielectric nanowire grids with minimal material embodiment render remarkable wideband polarization selectivity in both reflection and transmission.

A representative dielectric nanowire grid is illustrated in Fig. 1a along with its polarizing functionality. The structure is defined by the period $\Lambda$, wire fill factor $F$, height $h$, dielectric constant $\varepsilon_1$, and dielectric constant $\varepsilon_0$ of the host medium. Relative to the array geometry, transverse electric (TE) polarization corresponds to the electric field oscillating along the wire ($y$-axis) and the orthogonal polarization is denoted as transverse magnetic (TM) polarization. For ultra-sparse high-index nanowire grids where $F << 1$ and $\varepsilon_1 >> \varepsilon_0$, associated polarization selectivity arises in two fundamental aspects. First, there is a large difference in the diffraction strength between TE and TM polarizations. We denote the diffraction potential $V(q)$ for the $q$-th order diffraction process as the $q$-th Fourier-harmonic amplitude of the dielectric function $\varepsilon(x)$ for TE polarization and the reciprocal dielectric function $1/\varepsilon(x)$ for TM polarization [16]. These diffraction potentials can be approximated for high-index contrast ($\varepsilon_1 >> \varepsilon_0$) as

$$V_{\mathrm{TE}}(q) = (\varepsilon_1 - \varepsilon_0)F\cdot\mathrm{sinc}(Fq) \approx \varepsilon_1 F\cdot\mathrm{sinc}(Fq), \qquad (1)$$

$$V_{\mathrm{TM}}(q) = (1/\varepsilon_1 - 1/\varepsilon_0)F\cdot\mathrm{sinc}(Fq) \approx -(F/\varepsilon_0)\cdot\mathrm{sinc}(Fq), \qquad (2)$$

where $\mathrm{sinc}(Fq) = \sin(\pi Fq)/(\pi Fq)$. We note that the TM diffraction potential $V_{\mathrm{TM}}(q)$ is vanishingly small for $F << 1$ regardless of $\varepsilon_1$ whereas the TE diffraction potential $V_{\mathrm{TE}}(q)$ can be persistently high if $\varepsilon_1$ is large enough to maintain a sizable $\varepsilon_1 F$ product. Second, the array's average refractive index in this limit also supports the generation of polarization contrast. Using Rytov's effective medium theory [17], the zero-order effective refractive indices of the array for TE and TM polarizations are approximated as $n_{\mathrm{TE}} \approx$ $n_0(1+\varepsilon_1 F/\varepsilon_0)^{1/2}$ and $n_{TM} \approx n_0(1+F/2)$, respectively, where $n_0 = \varepsilon_0^{1/2}$. In these approximations, $n_{TM}$ approaches the background refractive index $n_0$ for $F << 1$ regardless of the $\varepsilon_1$ value while $n_{TE}$ is significantly larger than $n_0$ if the product $\varepsilon_1 F$ is comparable to or larger than $\varepsilon_0$. In this picture, TM-polarized input light is neither diffracted nor guided whereas TE-polarized light is both diffracted and guided and therefore capable of undergoing guided-mode resonance. The effects in play are enabled by the differences in the material manifestation of the device as experienced by the input light in alternate polarization states; this difference is conveniently and clearly expressible by the elementary diffraction potential and effective-medium theory as shown here. In the following analyses, we show how TE-polarized resonances and essential TM invisibility in a dielectric nanowire grid produce a broad reflection band with efficient polarization selectivity when the cross-sectional wire filling ratio $F$ tends toward an extremely small value.

Using rigorous coupled-wave analysis (RCWA) [16], we numerically calculate the zero-order reflectance ($R_0$) spectra under TE- and TM-polarized light incidence for three example designs with parameter sets ($\varepsilon_1$, $F$, $h/\Lambda$) = (100, 0.01, 0.315), (50, 0.02, 0.317), and (10, 0.1, 0.342). We take free-space ($\varepsilon_0 = 1$) as the host medium. In these examples, the product $\varepsilon_1 F$ is constant at 1 with wire height chosen to maximize the TE resonance reflectance. In Fig. 1b, the $R_0$(TE) spectra show broadband reflection over the normalized full-width at half-maximum bandwidth $\Delta\lambda/\Lambda \sim 20\%$. The normalized bandwidth of the reflection plateau where $R_0(\text{TE}) \geq 0.99$ is $\Delta\lambda/\Lambda \sim 13\%$ for all three cases. This broadband reflection occurs over a wide acceptance angle ~ 20° as shown in Fig. 2.

In contrast, as shown in Fig. 1b, the TM reflectance $R_0$(TM) is well below $10^{-2}$ and essentially negligible. Note that $R_0$(TM) is less than $3\times10^{-5}$ for $F = 0.01$ and $\varepsilon_1 = 100$. In more detail, $R_0$(TM) scales roughly with $2(n_{TM}-n_0)^2/(n_{TM}+n_0)^2 \approx F^2/4$ corresponding to the specular reflection from the homogeneous average effective film. Therefore, the polarization extinction ratio in reflection across the TE resonance bandwidth is approximately given by

$$R_0(\text{TE})/R_0(\text{TM}) \approx 4\varepsilon_1^2, \qquad (3)$$

with the condition $\varepsilon_1 F \approx 1$ to maintain an appreciable TE diffraction potential. The TE and TM field distributions at the center wavelength of the TE-reflection plateau for the examples with parameter sets ($\varepsilon_1$, $F$, $h/\Lambda$) = (100, 0.01, 0.315) and (10, 0.1, 0.342) are shown in Figs. 1c and 1d, respectively. In each case, there is a clear resonant field enhancement in the array for TE polarization whereas TM-polarized light propagates freely through the device without significant perturbation of the propagating wavefronts.

The devices presented herein operate under the guided-mode resonance effect. The broadband TE reflection is driven by resonant excitation and reradiation of lateral Bloch modes via the ±1 evanescent diffraction orders [18]. The generation of a wavevector directed along the +*z*-axis sustaining the propagation of the reflected wave is a diffractive effect and not related to reflections off grating ridge interfaces [19]. The nanogrids presented have exceedingly small fill factors and attendant thin grating ridges. They are capable of supporting only a single ridge mode. Thus, interference between multiple local ridge modes [20] plays no causal roles in these devices.

Considering the experimental feasibility of the proposed device concept, we note that various high-index materials are available to suit a given spectral region of interest. For example, semiconductors such as Si, GaAs, and Ge have dielectric constant in the range $\varepsilon = 10\sim20$ in the near-infrared and telecommunications bands [21]. For operation at longer wavelengths, much higher dielectric constants are available. $ZrSnTiO_3$ ceramics [22] and perovskite-related oxides [23] have $\varepsilon \sim 100$ in the THz domain. Artificial engineered materials are under development with hyperbolic metamaterials [24] for effective $\varepsilon \sim 100$ in the near-infrared domain and with H-shaped metallic patch arrays [25] for effective $\varepsilon \sim 1000$ at THz frequencies. Recalling Eq. (3), this value of dielectric constant implies a polarization extinction ratio ~ $4\times10^6$. Therefore, the proposed device class is promising to attain extreme polarization selectivity in various frequency domains.

We experimentally demonstrate a Si-nanowire-grid polarizing beam splitter in the near-infrared region. The fabrication steps include sputtering a 540-nm-thick amorphous Si film on a 1-mm-thick microscope slide glass, ultraviolet laser interference lithography to form a photoresist grating mask, reactive-ion etching using a $CHF_3$+$SF_6$ gas mixture, and post-etch $O_2$ ashing to remove residual photoresist. Figure 3a shows a photograph of nine fabricated devices on a 1×1 inch$^2$ substrate. Aiming for device operation in the telecommunications band over the 1300~1600 nm wavelength range, these devices have identical periods of $\Lambda$ = 854.0 nm but slightly different fill factors such that $F$ gradually decreases from 0.12 for the bottom-left device to 0.05 for the top-right device. Clearly visible is the semi-transparency in the device areas due to the low Si-wire filling fraction. The best performance is obtained with a device with $F$ = 0.1. Figures 3b, 3c, and 3d show cross-sectional and top-view scanning electron microscope (SEM) images of this device. The measured geometrical parameters are indicated therein. To keep the promising sample undestroyed, we took the cross-sectional micrographs in Fig. 3b from a sacrificial sample fabricated with an identical process while the top-view micrographs in Figs. 3c and 3d show the actual device whose optical spectrum is measured and presented here.

The fabricated sample is further prepared for spectral measurement. To establish an optically symmetric background environment, we place an index-matching fluid with refractive index 1.526 between the cover and substrate glass slides with refractive index 1.520. Thus the device is immersed in an approximately homogeneous host medium. Spectra are collected with an infrared spectrum analyzer (AQ6375, Yokogawa) and a supercontinuum light source (Koheras SuperK Compact, NKT Photonics). Figure 4 shows the measured TE and TM extinction ($1-T_0$) and reflectance ($R_0$) spectra in comparison with theoretical predictions. In the calculation, we apply the exact cross-sectional shape of the fabricated structure as shown in the insets of Figs. 4c and 4d. We also use the experimental dielectric constant $\varepsilon_1 = 12.25$ of our Si film that we determine with ellipsometry (VASE Ellipsometer, J. A. Woollam). Shown in Figs. 4a and 4b are the TE and TM extinction spectra in experiment and theory, respectively. We note that the extinction for lossless systems must be identical to the reflectance, i.e., $1-T_0 = R_0$, in the zero-order regime above the Rayleigh wavelength (white dotted lines). There is excellent quantitative agreement between theory and experiment across the wide angular and wavelength ranges considered. Figures 4c and 4d show the measured and theoretical spectra for the TE extinction, TE reflectance, and TM reflectance at normal incidence. The TE reflection bandwidth for $R_0$(TE) > 0.9 is ~190 nm in the experiment and ~200 nm in the theory. We attribute the higher TM reflectance in the experiment to the specular reflections at the cover and substrate glass surfaces where about 4% of the incident optical power is reflected from each. In the theoretical spectrum in Fig. 4d, the TM reflectance is below 0.5% over the wavelength region corresponding to the high TE-reflection plateau.

In conclusion, we present ultra-sparse dielectric nanowire grids that strongly polarize incident light in reflection and transmission across considerable spectral and angular extents. As supported by elementary effective-medium arguments, the nanogrid array is essentially invisible to TM polarized light while resonating effectively in TE polarization. In the context of conventional elements based on multilayer

Bragg reflection and multiple ridge mode interference, it is somewhat counterintuitive that a device with such minimal material manifestation can provide the striking effects observed. Hence, the proof-of-concept experimental demonstration provided is particularly relevant. It shows excellent quantitative agreement with theory. Our models therefore support pursuit of practical device design and implementation applying various available natural and engineered materials having high dielectric constants. Consequently, this device idea is feasible in wide spectral domains including the near-infrared, THz, and longer wavelength regions.

**Acknowledgements**

The research leading to these results was supported in part by the Texas Instruments Distinguished University Chair in Nanoelectronics endowment.

**Author contributions**

R.M. conceived the original idea and designed the initial devices. K.J.L. fabricated the prototype devices. J.W.Y. performed sample characterization, spectral measurements, and computations. J.W.Y and R.M. wrote the paper.

**Additional information**

Correspondence and requests for materials should be addressed to R.M. at magnusson@uta.edu.

**Competing financial interests**

The authors declare no competing financial interests.

## Figures and Figure Captions

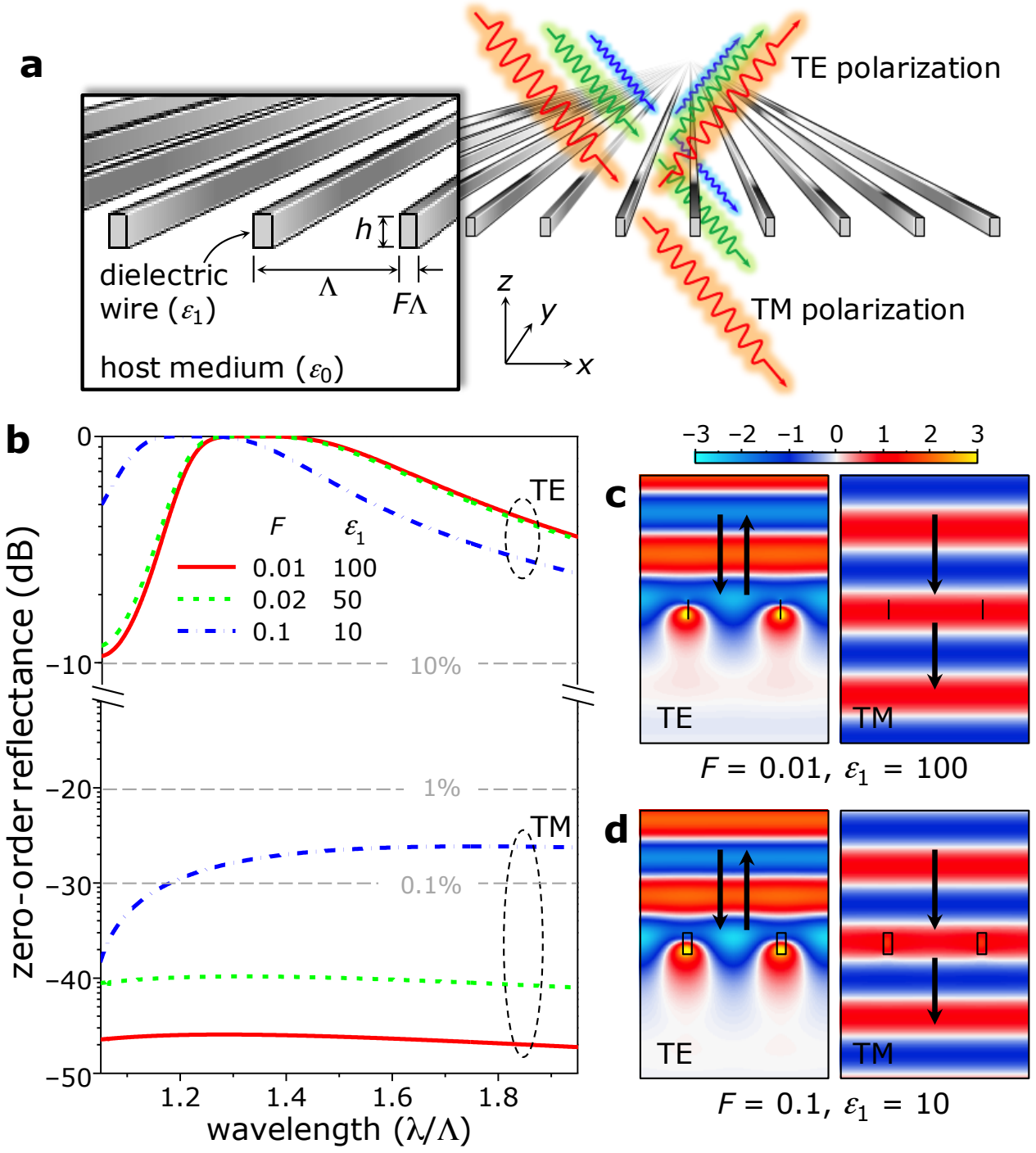

**Figure 1 | Theoretical performance of ultra-sparse dielectric nanowire arrays. a**, Schematic illustration of the proposed device. **b**, Zero-order TE and TM reflectance, $R_0$(TE) and $R_0$(TM), spectra under normal incidence for three different parameter sets ($\varepsilon_1$, $F$, $h/\Lambda$) = (100, 0.01, 0.315), (50, 0.02, 0.317), and (10, 0.1, 0.342). **c**, Field distributions under TE (left panel) and TM (right panel) polarized light incidence for the parameter set ($\varepsilon_1$, $F$, $h/\Lambda$) = (100, 0.01, 0.315) at wavelength $\lambda = 1.343\Lambda$. **d**, Field distributions under TE (left panel) and TM (right panel) polarized light incidence for the parameter set ($\varepsilon_1$, $F$, $h/\Lambda$) = (10, 0.1, 0.342) at wavelength $\lambda = 1.224\Lambda$. In **c** and **d**, the indicated fields are electric field for TE and magnetic field for TM. Their values are normalized by the incident field amplitude.

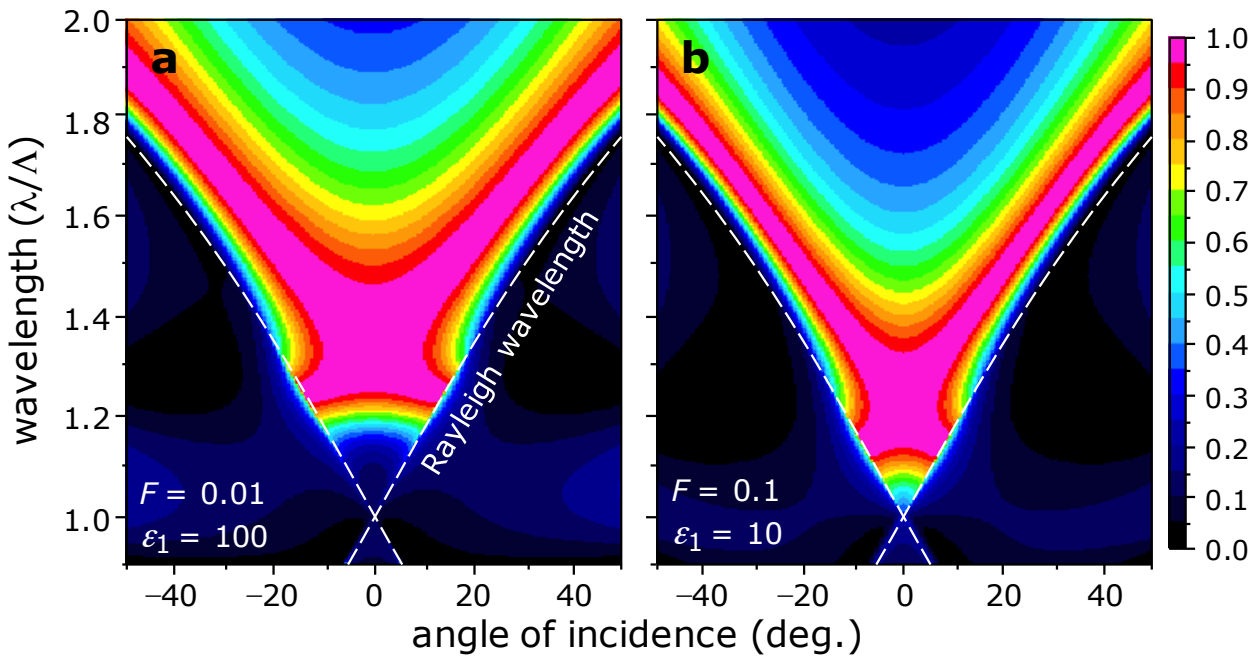

**Figure 2 | Angle-dependent TE-reflectance spectra for ultra-sparse dielectric nanowire designs with parameter sets:** **a**, ($\varepsilon_1$, $F$, $h/\Lambda$) = (100, 0.01, 0.315) and **b**, ($\varepsilon_1$, $F$, $h/\Lambda$) = (10, 0.1, 0.342).

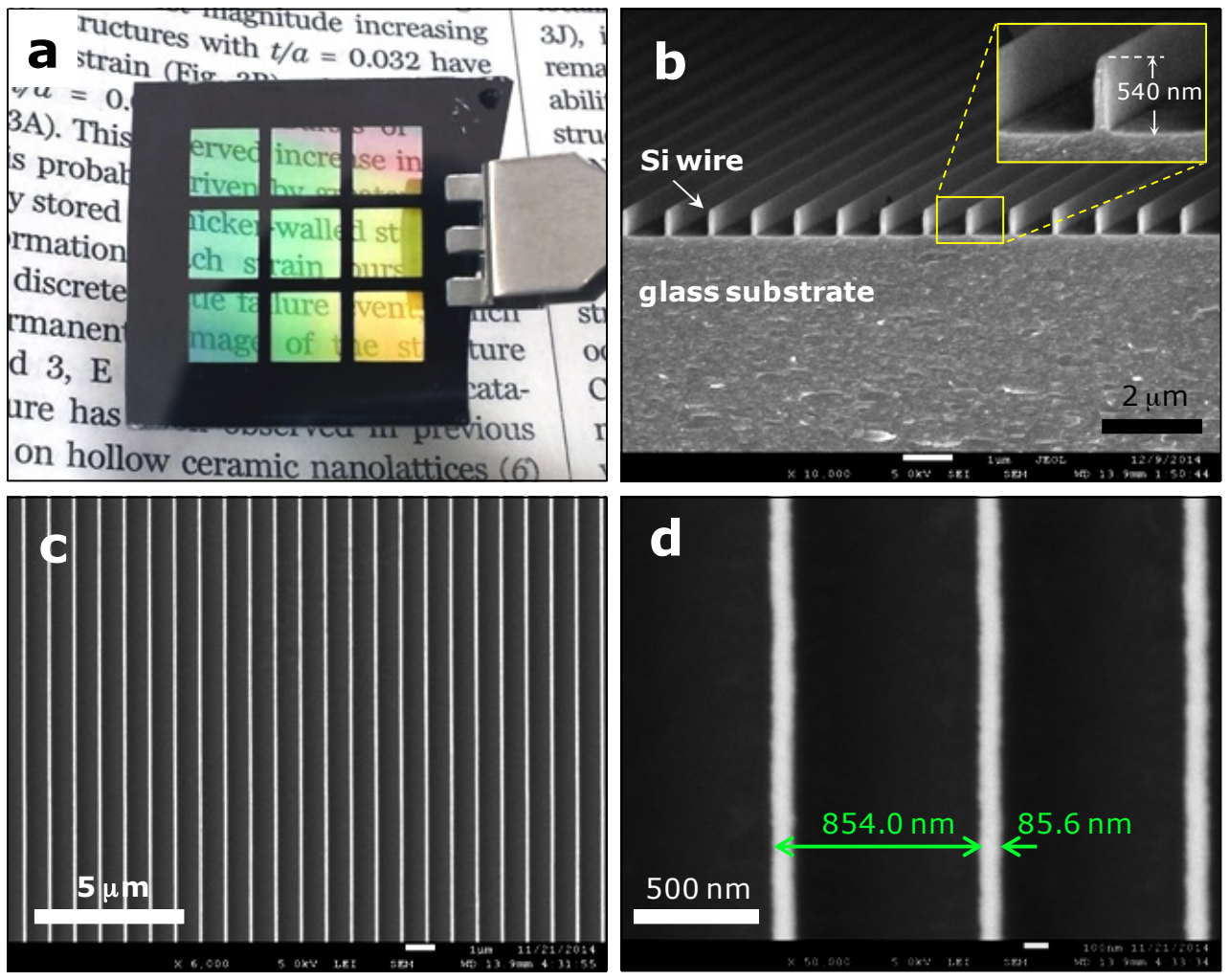

**Figure 3 | Device fabrication and characterization. a**, A photograph of nine ultra-sparse Si nanowire grids with different Si-wire fill factors on a 1×1 inch$^2$ glass substrate. Cross-sectional (**b**) and top-view (**c** and **d**) SEM images of the device.

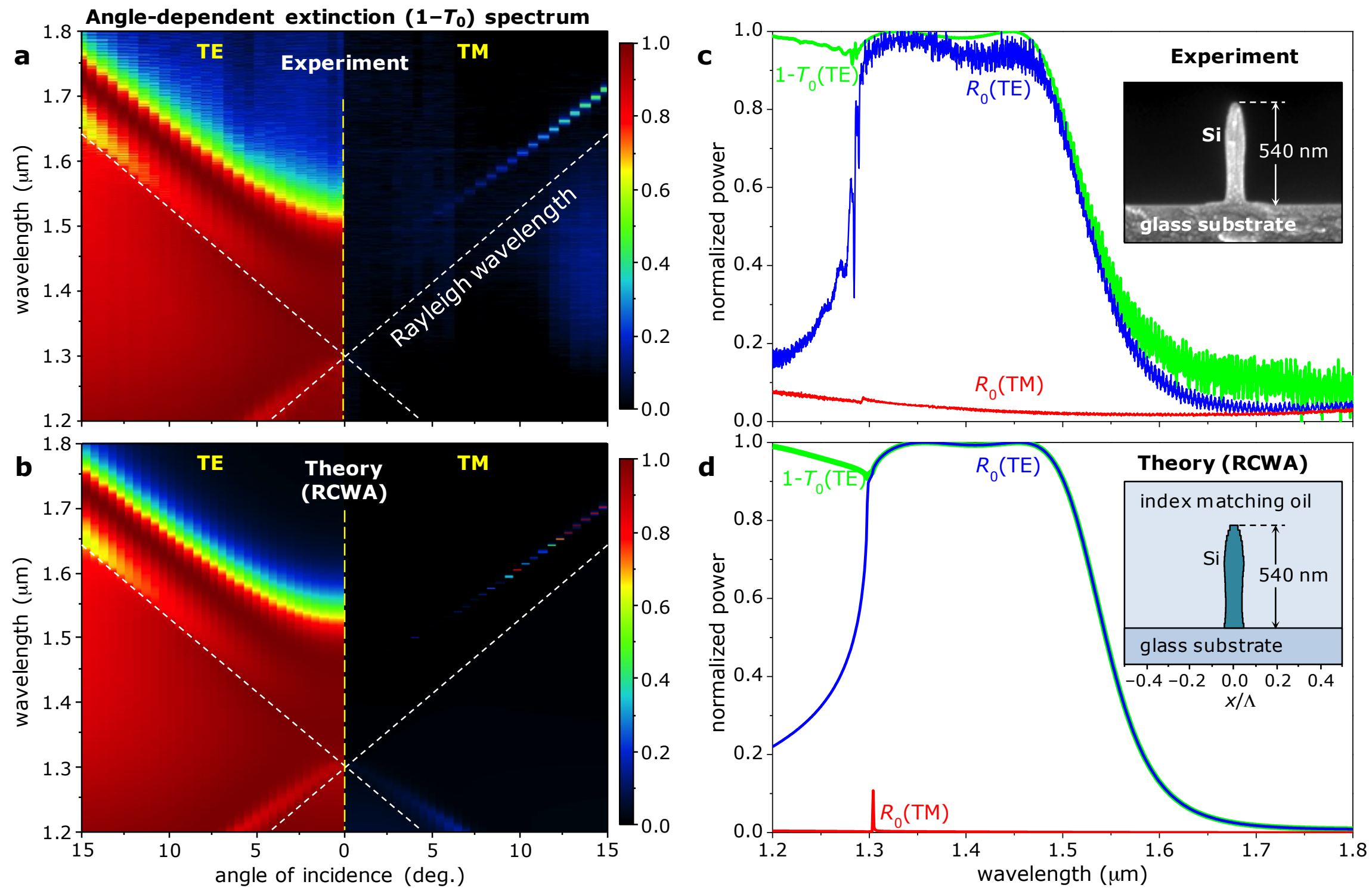

**Figure 4 | Experimental performance of an ultra-sparse Si nanowire array reflector/polarizer. a**, Measured angle-dependent extinction (1–$T_0$) spectra under TE- and TM-polarized light incidence. **b**, Calculated angle-dependent TE and TM extinction spectra for comparison. **c**, Measured spectra of the TE reflectance $R_0$(TE), TM reflectance $R_0$(TM), and TE extinction (1–$T_0$). **d**, Calculated spectra of the TE reflectance $R_0$(TE), TM reflectance $R_0$(TM), and TE extinction (1–$T_0$) for comparison.